\documentclass[12pt,preprint]{aastex}

\def\Ht{H_0^{-1}}
\def\locunit{{\rm km\;s^{-1}\,Mpc^{-1}}}
\def\zl{z_{\hbox{\tiny{L}}}}
\def\zs{z_{\hbox{\tiny{S}}}}
\def\reff{R_{\hbox{\tiny{eff}}}}

\begin{document}

\title{The Hubble time inferred from 10 time-delay lenses}

\author{Prasenjit Saha\altaffilmark{1,2}}
\author{Jonathan Coles\altaffilmark{1}}
\author{Andrea V. Macci\`o\altaffilmark{1}}
\author{Liliya L.R. Williams\altaffilmark{3}}

\altaffiltext{1}{Institute for Theoretical Physics, University of Z\"urich,
                 Winterthurerstrasse 190, 8057 Z\"urich, Switzerland}
\altaffiltext{2}{Astronomy Unit, Queen Mary and Westfield College,
                 University of London, London E1~4NS, UK}
\altaffiltext{3}{Department of Astronomy, University of Minnesota,
                 116 Church Street SE, Minneapolis, MN 55455}

\begin{abstract}
We present a simultaneous analysis of 10 galaxy lenses having
time-delay measurements.  For each lens we derive a detailed free-form
mass map, with uncertainties, and with the additional requirement of a
shared value of the Hubble parameter across all the lenses.  We test
the prior involved in the lens reconstruction against a
galaxy-formation simulation.  Assuming a concordance cosmology, we
obtain $\Ht=13.5^{+2.5}_{-1.3}\;\hbox{Gyr}$.
\end{abstract}


\section{Introduction}

If an object at cosmological distance is lensed into multiple images,
the light travel time for individual images differs.  For variable
sources, the differences are observable as time delays.  The delays
are of order
\begin{equation}
\Delta t \sim \frac{GM}{c^3} \sim (\Delta\theta)^2 \, \Ht
\label{delay-order}
\end{equation}
where $M$ is the lens mass and $\Delta\theta$ is the image separation
(in radians). As \cite{refsdal64} first pointed out, the effect
provides an independent way of measuring $\Ht$.
Time-delay measurements have made much progress over the past
decade and now at least 15 are available (details below).

While Eq.~(\ref{delay-order}) provides the order of magnitude, to
determine the precise factor relating time delays and $\Ht$ one has to
model the mass distribution.  An observed set of image positions,
rings, magnification ratios, and time delays is generically
reproducible by many different mass models.  This results in a large
model-dependent uncertainty on the inferred Hubble parameter, even
with perfect lensing data.  To appreciate how serious this
model-dependence is, compare the models of B0957+561 by
\cite{kundic97} and \cite{bernstein99}: the results are $H_0=64\pm13$
and $77^{+29}_{-24}\;\locunit$ respectively, both at $95\%$
confidence; the more general models in the latter paper yield {\em
larger\/} error-bars.  Alternatively, consider the nice summary in
Fig.~12 of \cite{courbin03} of published $H_0$ estimates and
uncertainties from individual lenses.  Among the lenses shown,
B1608+656 has all three of its independent time delays measured,
B1115+080 has two delays measured, whereas the others have one each.
One would expect these two best-measured lenses to be the best
constrained.  Yet B1608+656 has the largest error-bars on $H_0$ and
B1115+080 the second-largest.  This suggests that in the
less-constrained lenses the real uncertainties are much larger, but
have been underestimated because the fewness of constraints did not
force sufficient exploration of model-dependence.

A general strategy for dealing with the non-uniqueness problem is to
search through a large ensemble of models that can all reproduce the
observations \citep{ws00,oguri04,jakobsson05}.  In this paper we will
follow such a strategy, simultaneously modeling 10 time-delay lenses
coupled by a shared Hubble parameter.  The basic method is the same as
in \cite{sw04} and the accompanying {\em PixeLens\/} code, but a
number of refinements have been made.

\section{Modeling the lenses}

Table~\ref{tablobs} summarizes the lenses we have used.  By `type' we
mean the image morphology \citep{sw03}: AD = axial double, ID =
inclined double, SQ = short-axis quad, LQ = long-axis quad, IQ =
inclined quad. In B0957+561 two distinct source elements can be
identified, both are lensed into ID.

\begin{table}
\caption[]{Lenses and time delays}
\label{tablobs}
$$
\begin{array}{p{2cm}p{1cm}lll}
\hline\noalign{\smallskip}
Object     & type        & \zl & \zs & \Delta t \hbox{(days)} \\
\noalign{\smallskip}\hline\noalign{\smallskip}
J0911+055  & SQ          & 0.77 & 2.80  & 146\pm8^{\mathrm a}             \\
B1608+656  & IQ          & 0.63 & 1.39  & 32\pm2,5\pm2,40\pm2^{\mathrm b} \\
B1115+080  & IQ          & 0.31 & 1.72  & 13\pm2,11\pm2^{\mathrm c,d}     \\
B0957+561  & $2\times$ID & 0.36 & 1.41  & 423\pm1^{\mathrm e}             \\
B1104--181 & AD          & 0.73 & 2.32  & 161\pm7^{\mathrm f}             \\
B1520+530  & ID          & 0.71 & 1.86  & 130\pm3^{\mathrm g}             \\
B2149--274 & AD          & 0.49 & 2.03  & 103\pm12^{\mathrm h}            \\
B1600+434  & ID          & 0.42 & 1.59  & 51\pm4^{\mathrm i}              \\
J0951+263  & ID          & 0.24^{\mathrm j} & 1.24  & 16\pm2^{\mathrm j}  \\
B0218+357  & ID          & 0.68 & 0.96  & 10\pm1^{\mathrm k,l}            \\
\noalign{\smallskip}\hline
\end{array}
$$
{\rightskip 0pt plus 1cm
$^{\mathrm a}$\cite{hjorth02} \quad
$^{\mathrm b}$\cite{fassnacht02} \quad
$^{\mathrm c}$\cite{schech97} \quad
$^{\mathrm d}$\cite{barkana97} \quad
$^{\mathrm e}$\cite{oscoz01} \quad
$^{\mathrm f}$\cite{ofek03}  \quad
$^{\mathrm g}$\cite{burud02b}  \quad
$^{\mathrm h}$\cite{burud02a}  \quad
$^{\mathrm i}$\cite{burud00}  \quad
$^{\mathrm j}$\cite{jakobsson05} [photometric $\zl$] \quad
$^{\mathrm k}$\cite{biggs99} \quad
$^{\mathrm l}$\cite{cohen00}  \par}
\end{table}

We use {\em PixeLens\/} to generate an ensemble of 200 models.  Each
model in the ensemble consists of 10 pixelated mass maps and a shared
value of $\Ht$.  In addition to reproducing all the observed image
positions and time delays, the mass maps are required to satisfy a
prior.  Errors in the image positions and time delays are assumed
negligible, since they are much smaller than the range of models that
reproduce the data.  The details and justification of the prior are
given in Section~2 of \cite{sw04}, but basically the mass maps have to
be non-negative and centrally concentrated with a radial profile
steeper than $|\theta|^{-0.5}$, since the lenses are galaxies. With
one exception the mass maps are required to have $180^\circ$ rotation
symmetry; only B1608+656 is allowed to be asymmetric, because the lens
is known to contain two galaxies.  A constant external shear is
allowed for the lenses where the morphology show evidence of external
shear (all except B1608+656, B1104--181, B2149--274). The lensing
galaxies in B0957+561 and J0911+055 have cluster environments, but we
have not treated these lenses differently.  A concordance cosmology
with $\Omega_m=0.3,\Omega_\Lambda=0.7$ is assumed.

\begin{figure}
\epsscale{0.9}
\plotone{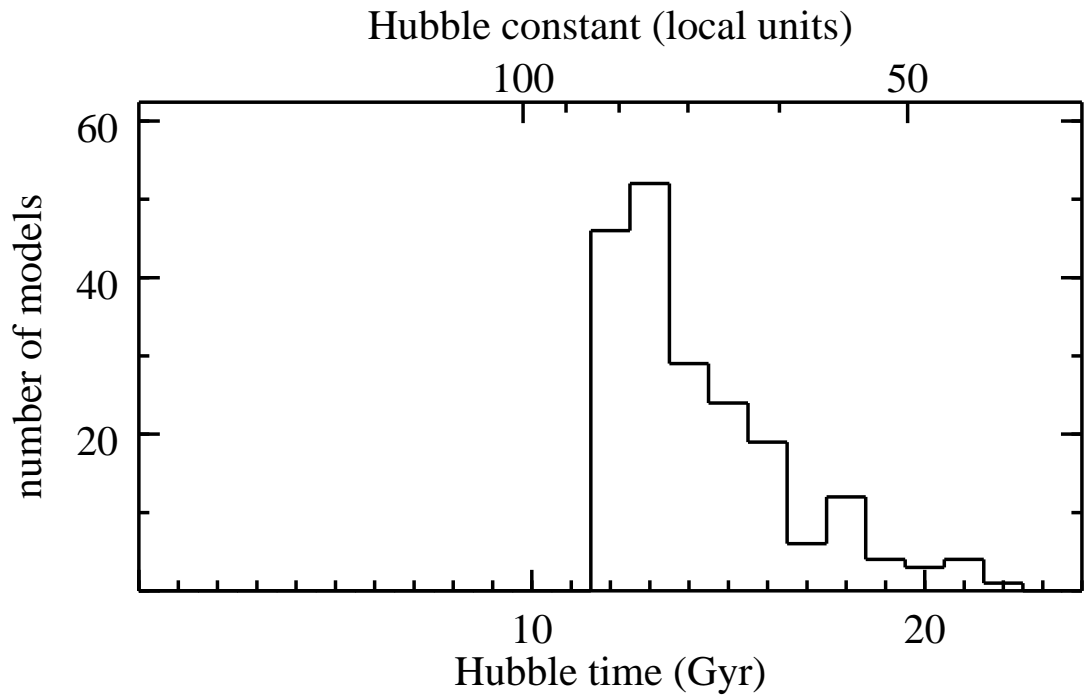}
\caption{Histogram of the ensemble of $\Ht$ values. The unbinned
distribution gives $\Ht=13.5^{+2.5}_{-1.2}\;$Gyr at 68\% confidence
and $13.5^{+5.6}_{-1.6}\;$Gyr at 90\% confidence.}
\label{pg}
\end{figure}

\begin{figure}
\epsscale{0.2}
\plotone{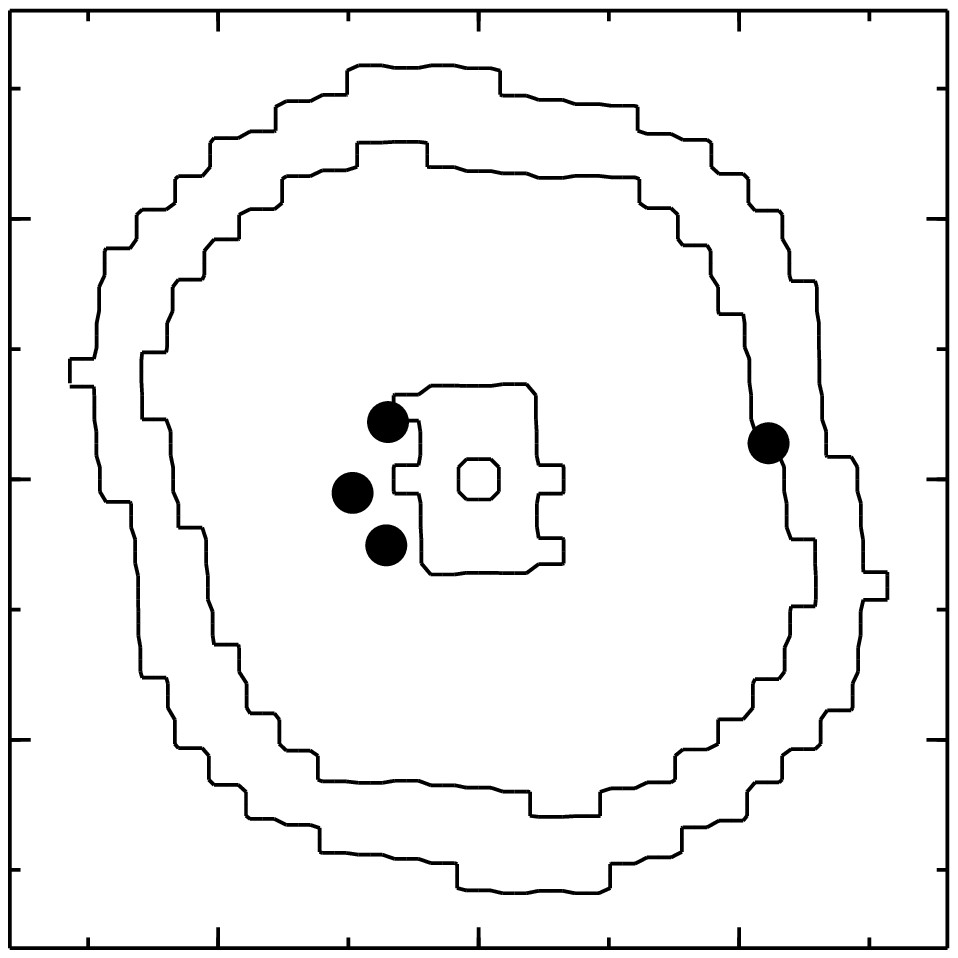}%
\plotone{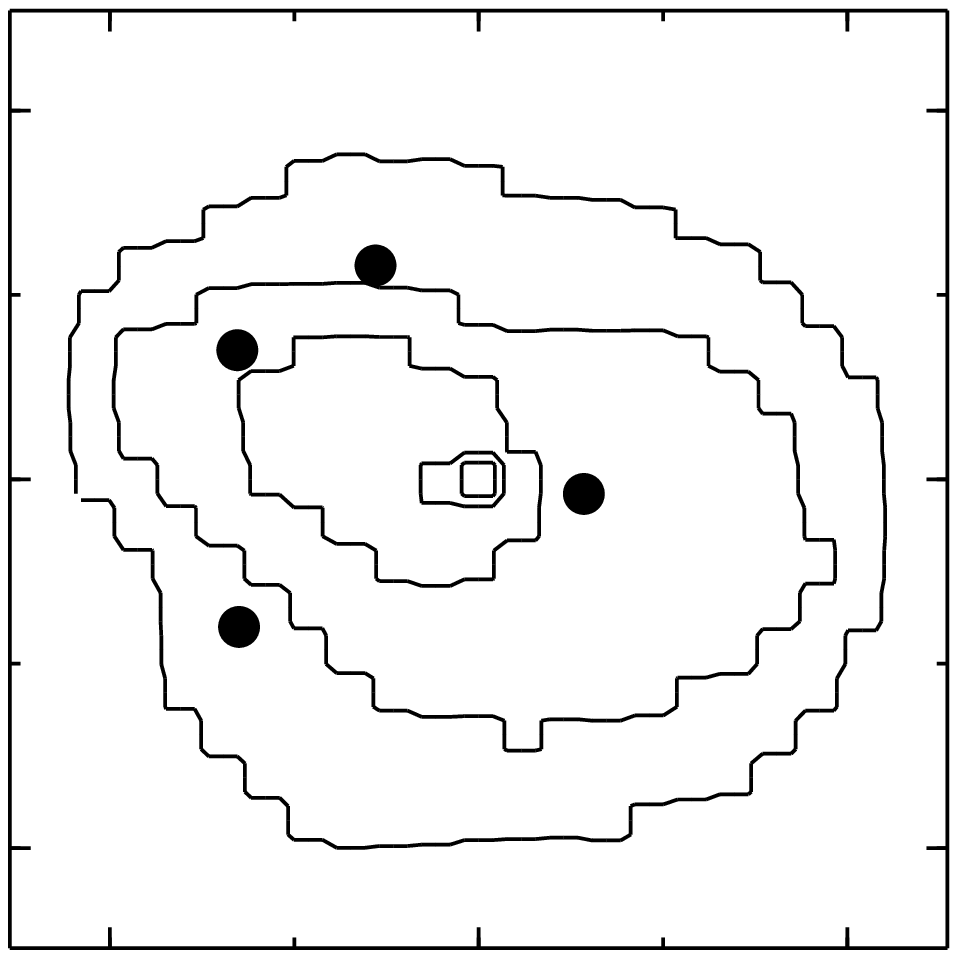}  \goodbreak
\plotone{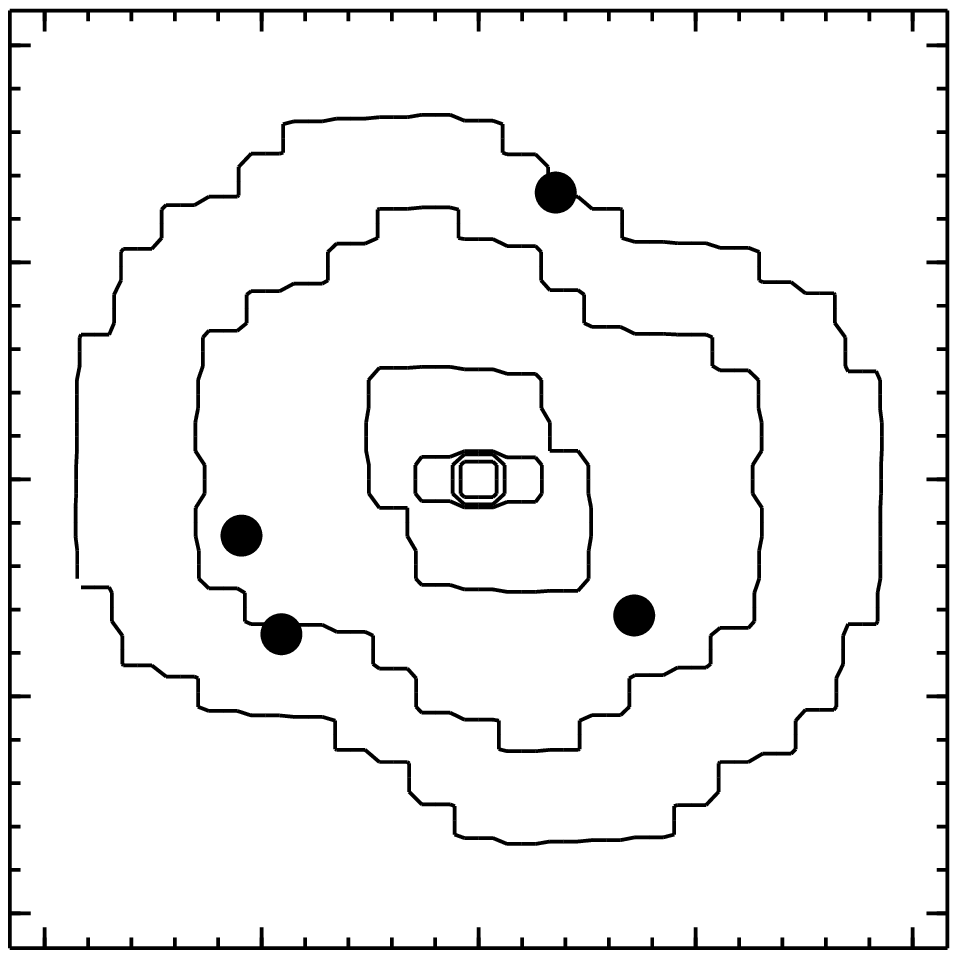}%
\plotone{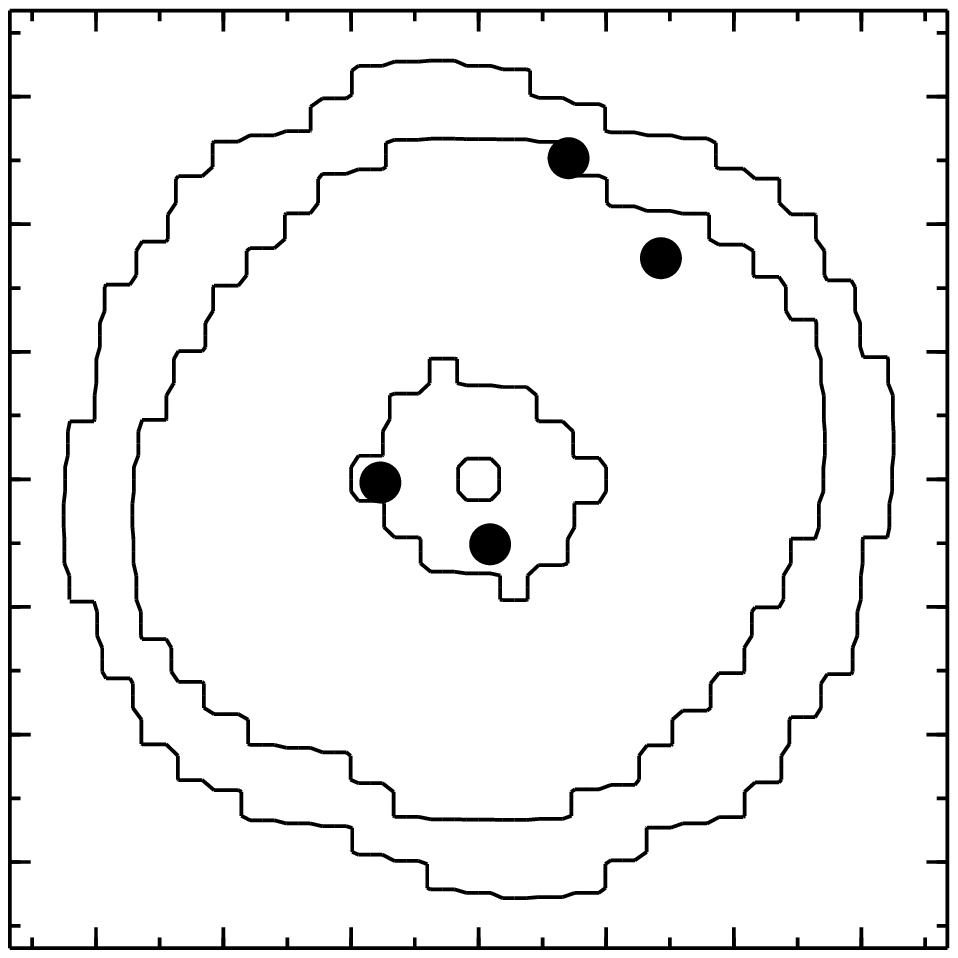}  \goodbreak
\plotone{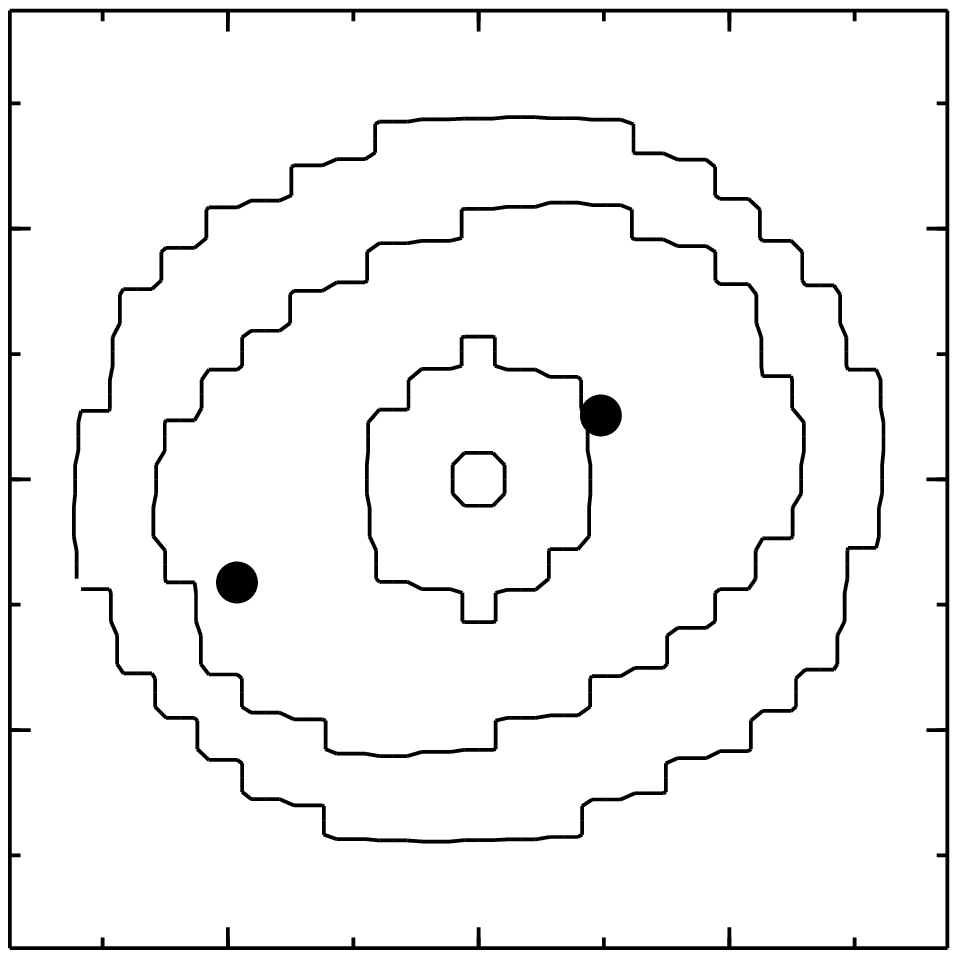}%
\plotone{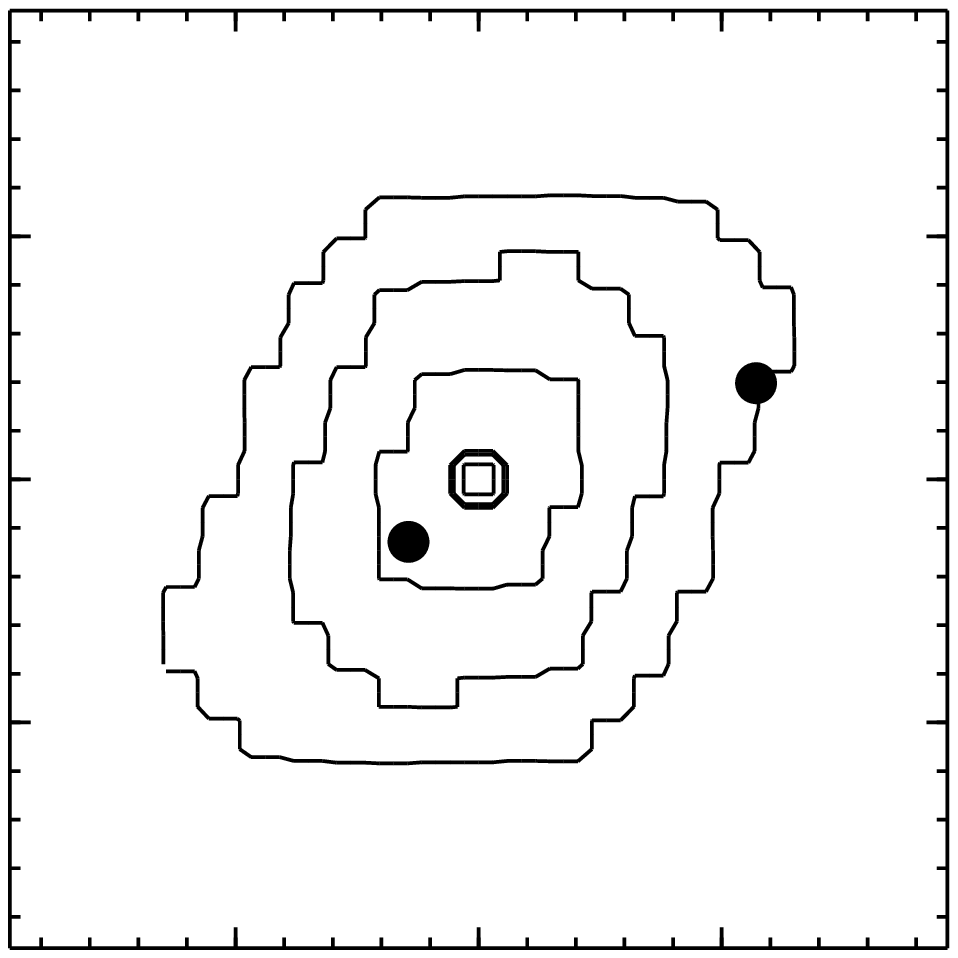}  \goodbreak
\plotone{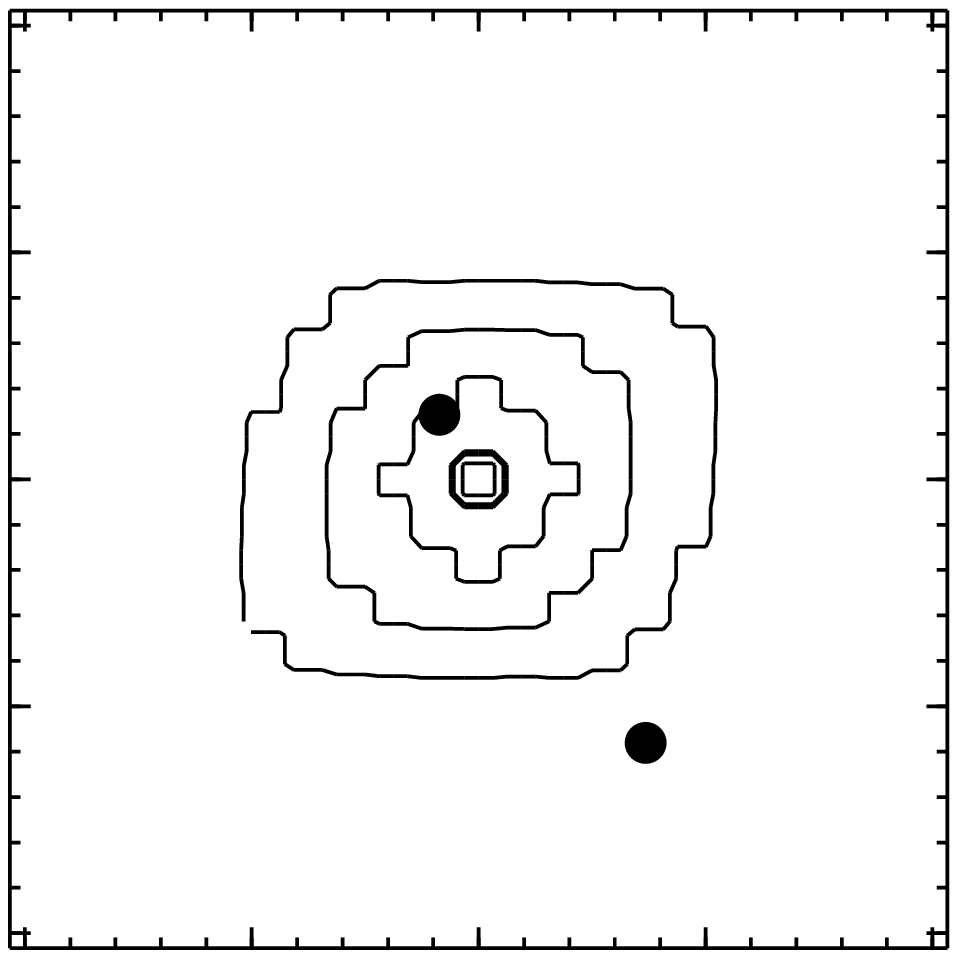}%
\plotone{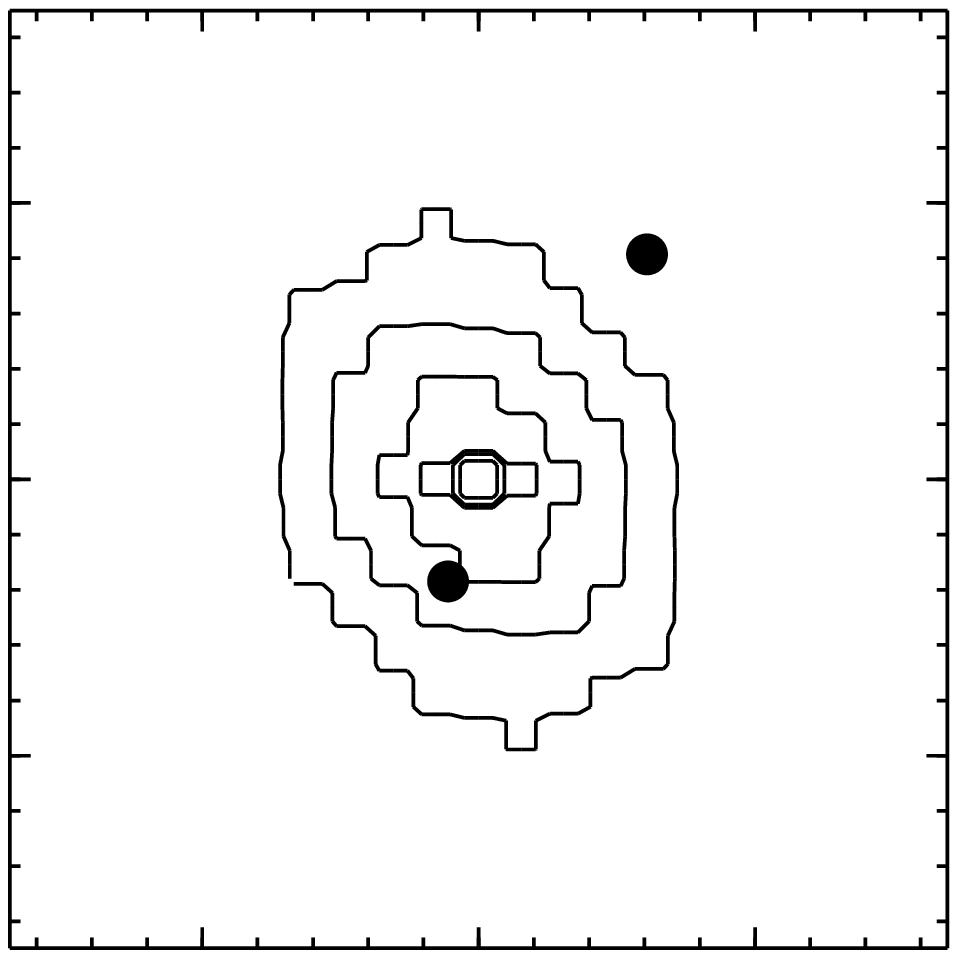}  \goodbreak
\plotone{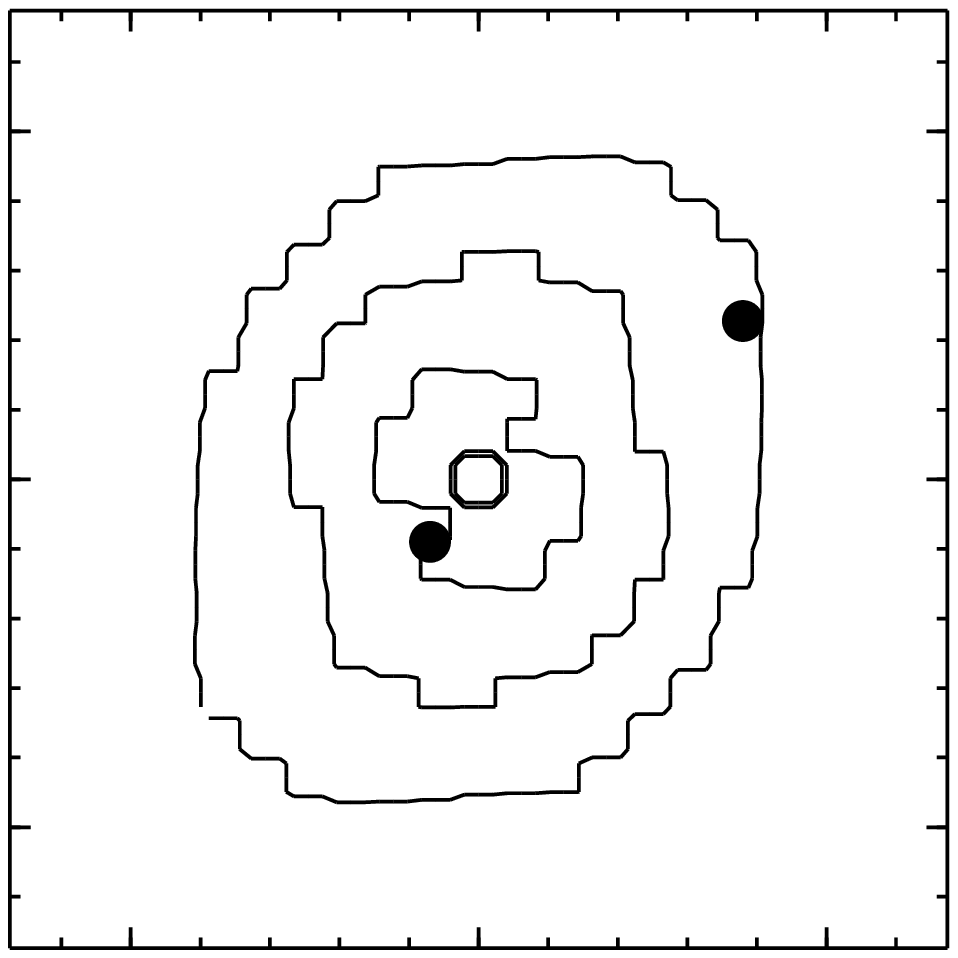}%
\plotone{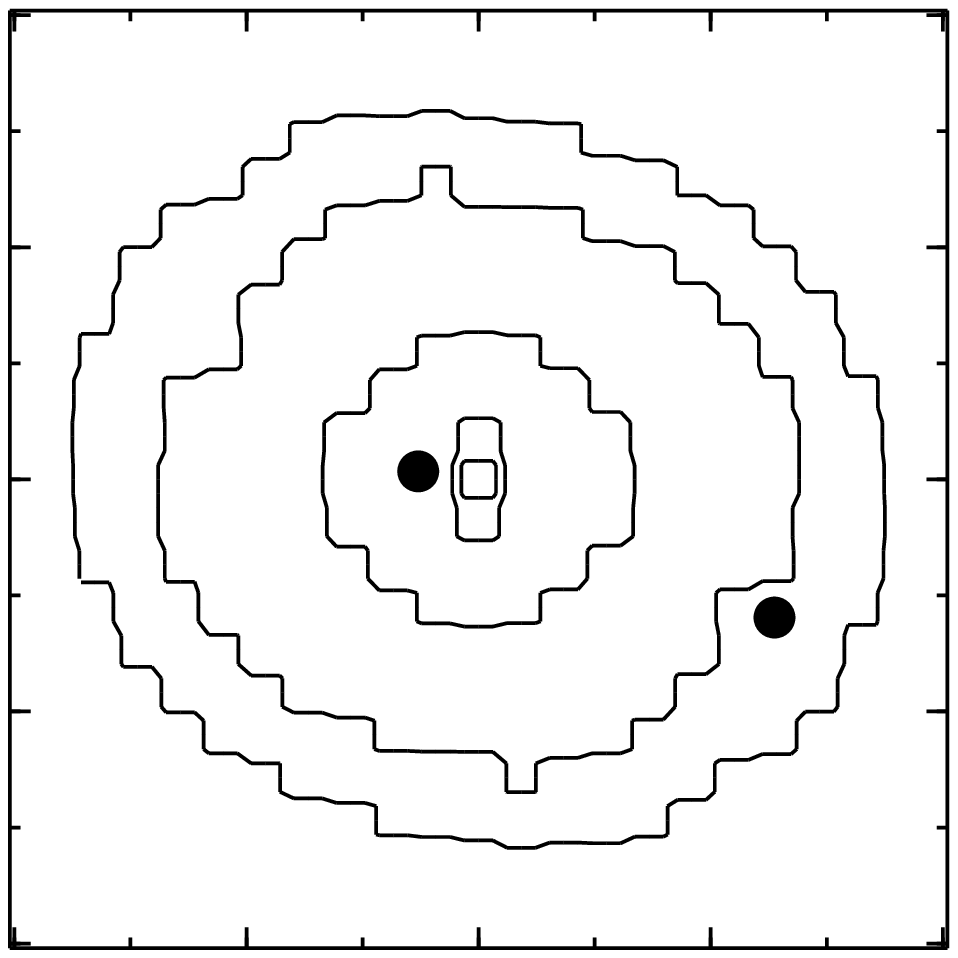}
\caption{Ensemble-average mass maps of the lenses: J0911+055 (upper
left), B1608+656 (upper right), B1115+080, B0957+561, B1104--181,
B1520+530, B2149--274, B1600+434, J0951+263, B0218+357.  The larger
tick marks in each panel correspond to $1''$. The contours are in
logarithmic steps, with critical density corresponding to the third
contour from the outside.}
\label{mass}
\end{figure}

We have not included magnification ratios as a constraint, for two
reasons: first, optical flux ratios may be contaminated by
microlensing \citep{keeton05} and differential extinction; second,
even tensor magnifications ---that is, relative magnifications along
different directions inferred from VLBI maps--- are very weakly
coupled with time delays \citep{rsw03}, because magnification measures
the local second-derivative of the arrival time. Stellar velocity
dispersions are available for some of the lenses, but we do not
attempt to incorporate them, because current methods for doing so
depend on strong assumptions about the mass distribution
\citep{koopmans06}.

There are five additional candidates we have postponed modeling.
B1830--211 has a time delay measurement \citep{lovell98} but the lens
position is uncertain \citep{courbin02,winn02}. B0909+532
\citep{ullan06} also has an uncertain galaxy position.  For B0435--122
\citep{kochanek06} and J1131-123 \citep{morgan06} our preliminary
modeling appeared to imply asymmetric lenses, whereas the image
morphologies suggest fairly symmetric lenses.  Finally, J1650+425 had
its time delay measured \citep{vuissoz06} as this paper was being
peer-reviewed.

We remark that while {\em PixeLens\/} in scientific terms is
essentially the same as in \cite{sw04}, it has undergone several
technical improvements.  The key parameter in the code's performance
is the total number of pixels (not pixels per lens) say $P$.  The
memory required scales as $P^2$ and the time scales as $P^3$.  The
maximum usable $P$ is in practice dictated not by time or memory but
by the accumulation of roundoff error.  Our earlier paper attempted
only 3 or 4 lenses at a time, going up to $P\simeq600$.  After
improving the control of roundoff error {\em PixeLens\/} can now go up
to $P\simeq2000$ and beyond without difficulty.  Meanwhile improving
the memory management and implementing multi-threading (which
parallelizes the computation if run on a shared-memory multi-processor
machine) and newer hardware have more than compensated for the $P^3$
increase in arithmetic.

We have previously done two different tests of the general method.  In
\cite{saha06} the algorithm is tested by feeding time delays sampled
from a model ensemble back into {\em PixeLens\/} and then recovering
the model $\Ht$.  This showed that any biases introduced by the
ensemble-generating process have affected $\Ht$ by less than 5\%, but
did not test the prior.  \cite{ws00} presented a blind test where one
author simulated data using simple model galaxies and a secret
fictional value of $H_0$, and the other author recovered that value
within uncertainties using an ancestor of {\em PixeLens}.  That
provided a basic test of the whole procedure, including the prior, but
still assumed that the models chosen by the first author for the test
were representative of real lensing galaxies.  A similar test using
current galaxy-formation simulations is desirable but technically
formidable; however, we carry out a simple version of such a test
below.

\section{Results}

Our $\Ht$ distribution is shown in Fig.~\ref{pg} and may be summarized
as
\begin{equation}
\Ht = 13.5^{+2.5}_{-1.2}\,\hbox{Gyr} \quad
(H_0 = 72^{+8}_{-11}\;\locunit)
\end{equation}
at 68\% confidence and $13.5^{+5.6}_{-1.6}\;$Gyr at 90\% confidence.
This estimate neglects measurement errors in the time delays. However,
we have verified by repeating the analysis with perturbed time delays
that the effect of measurement errors is very small.  Astrometric
errors are also very small.

Fig.~\ref{pg} is consistent with the analogous Figs.~8 and 11 in
\cite{sw04}, which derive from 2 time-delay quads and 4 doubles
considered separately.  But the constraints do not improve as much as
simple $1/\sqrt{N}$ would predict.  In fact, the uncertainties are far
from Gaussian, and some lensing configurations are much more useful
than others.  \cite{saha06} discuss this point in more detail and
conclude that a 5\% uncertainty on $\Ht$ is possible using 11 lenses,
provided the lenses all have favorable configurations.

Fig.~\ref{mass} shows ensemble average mass distributions for the 10
lenses.  Notice that some lenses, especially B1115+080, B1104--181,
and B1520+530, have twisting isodensity contours and/or
radially-dependent ellipticities, features that are not included in
parametrized models.

The lens galaxies have varying amounts of dark matter.  This follows
from \cite{ferreras05} who compare the total-mass profiles of 18
lensing galaxies, including 6 from the present sample, with
stellar-mass profiles from population-evolution models.  (The work
assumed $H_0^{-1}=14\;$Gyr which is well within our uncertainties, and
hence the results are valid for the models here.)  From their Table~1
we see that out to $\sim3\reff$, B1520+530 is mainly stars, B1115+080,
B1608+656, and B2149--274 have significant non-stellar mass, while
J0951+263 and B1104--181 are dominated by dark matter.

\begin{figure}
\epsscale{0.9}
\plotone{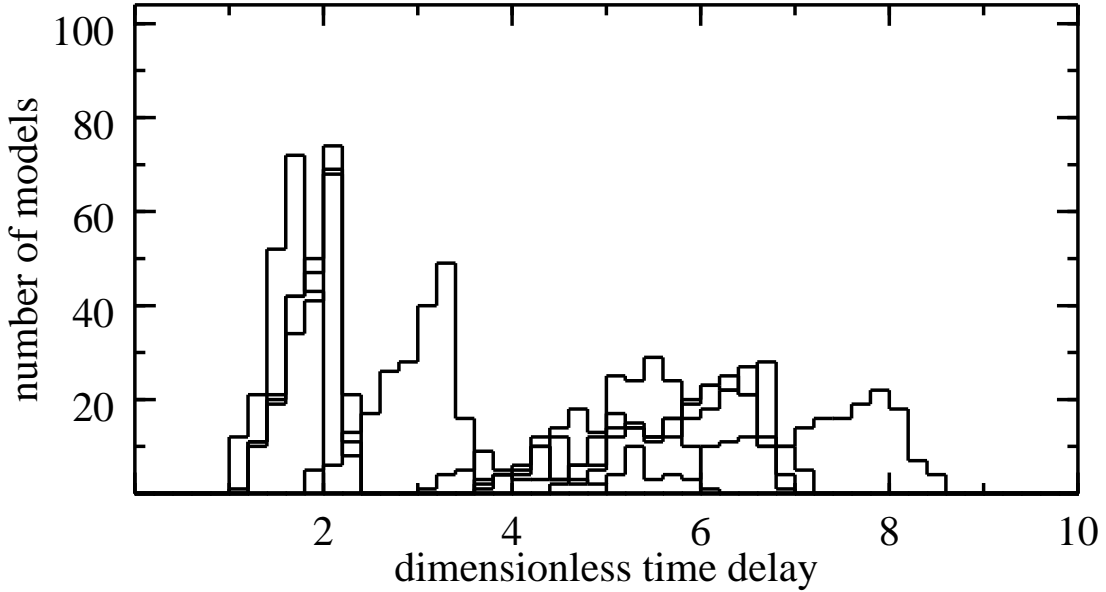}
\caption{Histograms of $\varphi$ for all ten lenses.
J0911+055, B1608+656, B1115+080 all peak around 2.
B0957+561 peaks around 3.
B1104--181 peaks around 2.
B1520+530, B1600+434, J0951+263, B0218+357 all peak around 6.
B2149--274 peaks around 8.}
\label{varphi}
\end{figure}

\begin{figure}
\epsscale{0.9}
\plotone{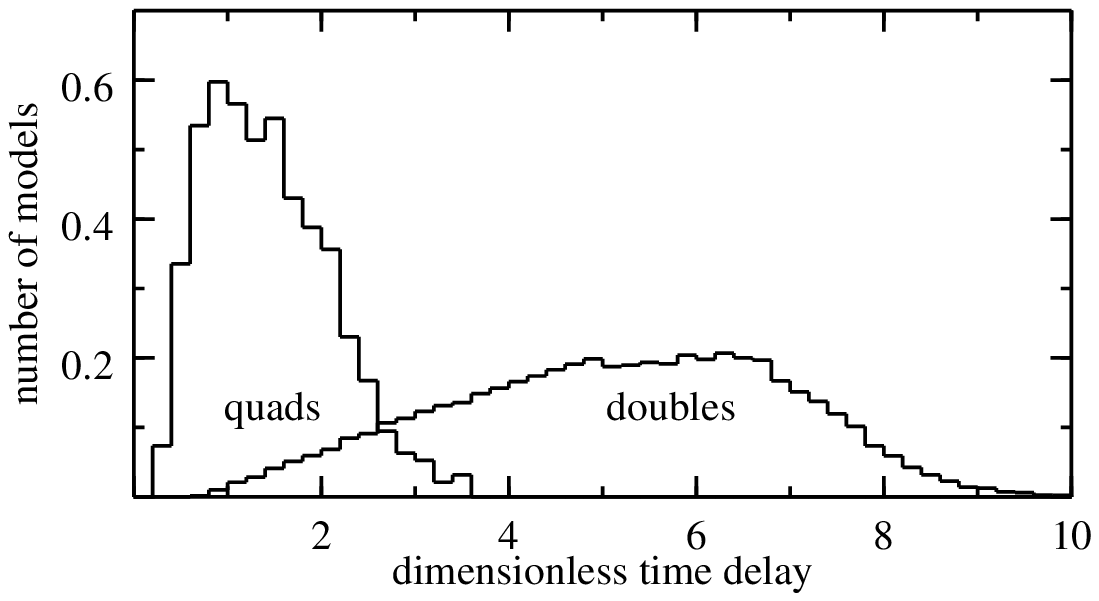}
\caption{Probability distribution of $\varphi$ for the simulated
galaxy. Doubles and quads are normalized separately.}
\label{andrea}
\end{figure}

\section{Lens models compared with a simulation}

We now address a simplified version of the question: are our lens
models typical of current galaxy-formation simulations?

The details of gas dynamics, star formation, AGN formation, and
feedback on galaxy scales are still uncertain.  With this caveat in
mind, we consider a single high resolution galaxy, extracted from an
$N$-body cosmological simulations and then resimulated using the
TreeSPH GASOLINE code \citep{wadsley04} including gas cooling, star
formation, supernova feedback and UV background.  The galaxy is an E1
or E2 triaxial elliptical with dominated by stars in the inner region
but overall $\sim80\%$ dark matter \citep{maccio06}.  Orienting this
galaxy randomly and ray-tracing with random sources \citep{maccio05}
we generated about 500 quads and 10000 doubles, and calculated time
delays for each of these.

As Eq.~(\ref{delay-order}) suggests, time delays generated from a
single galaxy will range over a factor of only a few, and cannot be
directly compared with the observed time delays, which range over a
factor of 40.  We therefore consider a dimensionless form of the
time-delay $\varphi$, given by
\begin{equation}
H_0 \Delta t = \varphi \, {\textstyle\frac1{16}} (\theta_1+\theta_2)^2 D
\end{equation}
where $\theta_1,\theta_2$ are the lens-centric distances (in radians)
of the first and last images to arrive, $\Delta t$ is the observed
time delay between them, and $D$ is the usual distance factor in
lensing.  This factors out the dependence of the time delay on
cosmology (through $H_0$ and $D$) and on the scale of the lens
(through $[\theta_1+\theta_2]^2$), leaving $\varphi$ dependent only on
the shape and steepness of the lens and on the source position with
respect to the caustics \citep{saha04}.

Fig.~\ref{varphi} shows the histograms of $\varphi$ in our lens-model
ensembles.  The quads all peak around 2, while the doubles mostly peak
around 5--8; the exceptions are B0957+561 peaking around 3 and
B1104--181 peaking around 2.  Since B0957+561 is in a cluster, it is
plausible that the mass profile is unusually shallow, thus reducing
the time delay through the well-known steepness degeneracy. The low
value for B1104--181 is more puzzling.
Fig.~\ref{andrea} is simpler, showing the probability distributions of
$\varphi$ for doubles and quads generated by the single simulated
galaxy.

Figs.~\ref{varphi} and \ref{andrea} are not quite equivalent, but we
can think of both as derived from an underlying
${\rm prob}(\varphi | \hbox{galaxy, lensing obs})$.
Each histogram in Fig.~\ref{varphi} weights 
this probability distribution by
observation selection effects and by the {\em PixeLens\/} prior, and
then marginalizes over galaxies while holding the lensing observables
fixed. Fig.~\ref{andrea} marginalizes over lensing observables
(separately for doubles and quads) while holding the galaxy fixed.
Bearing this difference in mind, the simulated galaxy appears typical
of our lens models.  The most noticeable difference is the absence of
observed quads with $\varphi$ close to zero; but that is an expected
observational selection effect, because very short time delays are
unlikely to be measured.

We conclude that the {\em PixeLens\/} prior, as far as this
preliminary experiment can reveal, is consistent with galaxy-formation
simulations. Further comparisons with simulated galaxies and
fine-tuning of the prior are desirable in future work.

\section{Discussion}

We have expressed our main result (Fig.~\ref{pg}) preferentially in
terms of $\Ht$ rather than $H_0$ because the former appears more
naturally in lensing theory.  But it is interesting to continue with
$\Ht$ in comparing with other techniques, because $\Ht$ has a simple
interpretation quite generally: it is $a/\dot a$ or the doubling-time
for metric distances at the current expansion rate.
Coincidentally,\footnote{Though a spoof paper by
D.~Scott ({\tt astro-ph/0604011}) develops a conspiracy theory for
this.} in the concordance cosmology ($K=0,\Omega_m\simeq\frac14,w=-1$)
$\Ht$ also equals the expansion age of the universe, within
uncertainties.  In particular, $\Ht$ estimates can be immediately
compared with globular-cluster ages, such as in \cite{krauss03}.

The well-known recent measurements of $\Ht$, expressed in Gyr are:
\begin{enumerate}
\item $13.6\pm1.5$ from \cite{freedman01}, who combine several
different indicators calibrated using Cepheids;
\item $15.7\pm0.3\;\hbox{(statistical)}\pm1.2\;\hbox{(systematic)}$
from \cite{sandage06}, using SN\thinspace Ia distances calibrated using
Cepheids;
\item $13.6\pm0.6$ from \cite{spergel06}, using the CMB fluctuation
spectrum.
\end{enumerate}
Our result is consistent with any of these.

It is worth noting, however, that the Hubble parameter appears in very
different guises in different techniques.  The distance-ladder methods
measure the local cosmological expansion rate, independent of the
global geometry.  By contrast, in the CMB, $\Ht$ is one parameter in a
global cosmological model.  Lensing is different again: here one
assumes a global geometry and then measures a single scale parameter.
The same is true of Sunyaev-Zel'dovich and X-ray clusters.  The latter
technique has made significant progress recently \citep{jones05} but
thus far still relies on strong assumptions: spherical symmetry of the
cluster potential and hydrostatic equilibrium of the gas.  In
principle, lensing time delays can determine the global geometry as
well \citep{refsdal66} but the amount of data needed is not
observationally viable yet.

Whether lensing time delays can get the uncertainties in the Hubble
parameter down to the 5\% level is an open question.  Maybe
galaxy-lens models can be constrained enough to determine $\Ht$ to
better than 5\%, thus making lensing the preferred method
\citep{schech04}; or maybe the approach is best used in reverse,
inputting $\Ht$ to constrain galaxy structure \citep{kochanek06}.
Fortunately, either outcome is worthwhile, and the basic technique
will be the same. So whether the optimists or the pessimists are
right, the usual cliches of ``more data!''  [time-delay measurements]
and ``more theory!''  [lens models] are both apt.


\end{document}